\documentclass[preprint,12pt]{elsarticle}
\usepackage{amssymb}
\usepackage{amsthm}
\usepackage{lineno}
\usepackage{tikz}
\usetikzlibrary{shapes, arrows}
\usepackage{amssymb,amsmath}
\usepackage{graphicx}
\usepackage{multirow}
\usepackage{hyperref}
\usepackage{appendix}
\usepackage{mathtools}
\usepackage{xcolor}
\usepackage{siunitx}
\usepackage{caption}
\usepackage{graphicx}% Include figure files
\usepackage{dcolumn}% Align table columns on decimal point
\usepackage{bm}

\journal{Journal of Computational Science}

\begin{document}

\begin{frontmatter}

\title{Material Hardness Descriptor Derived by Symbolic Regression}

\author[inst1,inst2,inst3]{Christian Tantardini} 
%[orcid=0000-0002-2412-9859]
\ead{christiantantardini@ymail.com}

\affiliation[inst1]{organization={Hylleraas center, Department of Chemistry, UiT The Arctic University of Norway},
            addressline={P.O. Box 6050 Langnes}, 
            city={Troms\o},
            postcode={N-9037},
            country={Norway}}

\affiliation[inst2]{organization={Department of Materials Science and Nanoengineering, Rice University},
            addressline={6100 Main St}, 
            city={Houston},
            postcode={77005}, 
            state={Texas},
            country={United States of America}}

\affiliation[inst3]{organization={Institute of Solid State Chemistry and Mechanochemistry SB RAS},
            addressline={Kutateladze 18}, 
            city={Novosibirsk},
            postcode={630128}, 
            country={Russian Federation}}

\author[inst4]{Hayk A. Zakaryan} 
%[orcid = 0000-0002-6105-3013]

\affiliation[inst4]{organization={Yerevan State University},
            addressline={1 Alex Manoogian St.}, 
            city={Yerevan},
            postcode={0025}, 
            country={Armenia}}

\author[inst5]{Zhong-Kang Han} 
%[orcid=0000-0003-1489-6824]

\affiliation[inst5]{organization={Zhejiang University},
            addressline={866 Yuhangtang Rd}, 
            city={Hangzhou},
            postcode={310027}, 
            country={China}}

\author[inst6]{Tariq Altalhi} 
%[orcid = 0000-0002-5470-8579]

\affiliation[inst6]{organization={Chemistry Department, Taif University},
            addressline={P.O. Box 11099}, 
            city={Taif},
            postcode={21944}, 
            country={Saudi Arabia}}

\author[inst5]{Sergey V. Levchenko} 
%[orcid = 0000-0001-5813-8473]

\affiliation[inst7]{organization={Skolkovo Institute of Science and Technology},%Department and Organization
            addressline={Bolshoy Boulevard 30, bld. 1}, 
            city={Moscow},
            postcode={121205}, 
            country={Russian Federation}}

\author[inst7]{Alexander G. Kvashnin} 
%[orcid =0000-0002-0718-6691]

\author[inst2,inst6]{Boris I. Yakobson}  
%[orcid = 0000-0001-8369-3567]
\ead{biy@rice.edu}

\begin{abstract}
Hardness is a materials' property with implications in several industrial fields, including oil and gas, manufacturing, and others.
However, the relationship between this macroscale property and atomic (i.e., microscale) properties is unknown and in the last decade several models have unsuccessfully tried to correlate them in a wide range of chemical space.
The understanding of such relationship is of fundamental importance for discovery of harder materials with specific characteristics to be employed in a wide range of fields.
In this work, we have found a physical descriptor for Vickers hardness using a symbolic-regression artificial-intelligence approach based on compressed sensing.
SISSO (Sure Independence Screening plus Sparsifying Operator) is an artificial-intelligence algorithm used for discovering simple and interpretable predictive models.
It performs feature selection from up to billions of candidates obtained from several primary features by applying a set of mathematical operators.
%In our work the primary features are obtained from  atomistic simulations for 61 materials.
The resulting sparse SISSO model accurately describes the target property (i.e., Vickers hardness) with minimal complexity.
We have considered the experimental values of hardness for binary, ternary, and quaternary transition-metal borides, carbides, nitrides, carbonitrides, carboborides, and boronitrides of 61 materials, on which the fitting was performed..
The found descriptor is a non-linear function of the microscopic properties, with the most significant contribution being from a combination of Voigt-averaged bulk modulus, Poisson's ratio, and Reuss-averaged shear modulus.
Results of high-throughput screening of 635 candidate materials using the found descriptor suggest the enhancement of material's hardness through mixing with harder yet metastable structures (e.g., metastable VN, TaN, ReN$_2$, Cr$_3$N$_4$, and ZrB$_6$ all exhibit high hardness).
\end{abstract}


\end{frontmatter}


\section{Introduction}
\label{sec:intro}

\begin{figure}[h!]
\centering
\begin{tikzpicture}[node distance=2cm, auto]
    % Define block styles
    \tikzstyle{block} = [rectangle, draw, fill=blue!20, 
        text width=14em, text centered, rounded corners, minimum height=4em]
    \tikzstyle{line} = [draw, -latex']

    % Place nodes
    \node [block] (training) {Training set of materials (61 materials)};
    \node [block, below of=training] (primary) {Primary features (first-principles calculations)};
    \node [block, right of=primary, node distance=8cm] (target) {Target property (experimental values of Vicker's hardness)};
    \node [block, below of=primary, node distance=3cm] (complex) {Complex features (combinations of primary features by a set of mathematical operators)};
    \node [block, below of=complex] (sisso) {Feature selection and model construction (SISSO)};
    \node [block, below of=sisso] (descriptor) {Physically interpretable microscopic descriptor};
    \node [block, below of=descriptor] (screening) {High-throughput screening (635 materials with different composition and crystal structure)};

    % Draw edges
    \path [line] (training) -- (primary);
    \path [line] (primary) -- (complex);
    %\path [line] (primary) -- (sisso);
    \path [line] (target) -- (sisso);
    \path [line] (complex) -- (sisso);
    \path [line] (sisso) -- (descriptor);
    \path [line] (descriptor) -- (screening);
\end{tikzpicture}
\caption{Schematic workflow of the SISSO method, illustrating the steps from the training set and primary features to the final high-throughput screening.}
\label{fig:scheme}
\end{figure}

Hardness is a mechanical property of materials important for several industrial applications.
In particular, hardness is measured and used as a parameter determining the type of application itself for materials in construction or manufacturing, e.g., cutting, drilling, or grinding \cite{kanyanta_hard_2016,kasonde_future_2016,haines_synthesis_2001}.
Over the years, various scales of hardness have been proposed. 
The Vickers scale is considered universal because it spans both macro- and micro-scales and is independent of the size of the indenter \cite{broitman_indentation_2016}. 
Thus, Vickers hardness is commonly measured in various applications to determine whether a material is superhard, typically 40 GPa or higher hardness \cite{solozhenko_synthesis_2005, kaner_designing_2005}. 
In some applications these materials are required to fulfil additional requirements.
For example, they need to preserve their hardness at high pressure and temperature, be non-toxic, and so on. 
Therefore, searching for hard and superhard materials with different chemical compositions remains an important challenge.
The best candidates for superhard materials are borides, carbides, and nitrides of metals\cite{solozhenko_ultimate_2009, solozhenko_mechanical_2001}.

In order to find materials with high hardness among many candidates, one can synthesize and test all of them one by one. 
However, this is obviously a very inefficient approach. Alternatively, one can find a correlation between hardness and features (or their mathematical combinations) that are easy to evaluate.
This correlation can then be used to quickly explore the chemical space of candidate materials (see also discussion and Fig.3 in Penev \textit{et al.} \cite{Penev_2021}). Such a combination of features is called descriptor. 

Both macroscopic properties (fluidity, elastic stiffness, ductility, strength, crack resistance and viscosity) and microscopic  properties (from atomistic simulations) can be important constituents of the descriptor \cite{broitman_indentation_2016}.
Which features are most important or sufficient is unknown, but it is known that none of the single features tested so far is a good descriptor.
On the other hand, a set of physically relevant features can be relatively easily obtained from atomistic simulations.
Therefore, it is attractive to explore whether a mathematical combination of such features (perhaps non-linear) correlates with Vickers hardness.
In the last decade several semi-empirical models were developed that use elastic properties as input to predict the hardness \cite{teter_computational_1998, chen_modeling_2011, mazhnik_model_2019}. 
Low accuracy of such models points out that other features must be included.

Recently, Podryabinkin \textit{et al.} \cite{podryabinkin_nanohardness_2022} proposed an alternative approach to calculate the hardness, which utilises first-principles calculations and machine learning potentials that were actively learned on local atomic environments in order to explicitly model the process of nanoindentation.
Although this method is highly accurate, it is too computationally expensive for a high-throughput screening.
Thus, there is a need for a physical, easily computable descriptor of hardness that can be used for a high-throughput search for superhard materials in the vast chemical space.  
Such descriptor is found here using sure independence screening plus sparsifying operator (SISSO) \cite{ouyang_sisso_2018,Han2021-so,ouyang2019simultaneous,purcell2023recent,xu2022sisso,wei2022sisso}.
In the past, neural networks have been used several times to describe Vickers hardness \cite{pourasiabi2012development,sembiring2020artificial,abd2020simulation,vermeulen1996prediction}.
However, the considered chemical and structural spaces were quite limited.
Moreover, no microscopic information was used in these studies.
Also, due to the nature of such machine learning methods, all physical meaning is lost, providing no insight into the structural features that could be explored to create new, harder materials.

\section{Methods}
\label{sec:methods}

The workflow that we have used and showed in Fig.~\ref{fig:scheme} is described below.
The initial step is to define the set of primary features of materials.
The primary features are readily available or obtainable properties that may have a physical relation to the target property (i.e. Vicker's hardness) of 61 compounds for which the experimental value is known. 
We have included all features that can be obtained from atomistic simulations easier than hardness itself and are physically related to bond strength, bond breaking, and restructuring \cite{broitman_indentation_2016}.
For example, first ionization energy is related to chemical bond hardness \cite{Shankar_2009}. 
Bulk and shear modules are used in the well-known Chen's model \cite{chen_modeling_2011}. 
Other features are mechanical properties (Young's modulus, Poisson's ratio, and so on) that are related to hardness, but this relationship is non-trivial \cite{broitman_indentation_2016}.
We use SISSO to both select most relevant features (or rather combination of features) from the initial set and find a (possibly non-linear) relationship between these features and hardness, as described below.
The accuracy of the model and in particular its predictive power (accuracy of predicting hardness for materials not included in the training set) indicate whether the initial set of primary features is sufficient for predicting hardness.

Primary features are combined using a set of mathematical operators to form a large number (up to tens of billions) of complex features.
Each complex feature is a (generally) non-linear formula including one, two, or more primary features, depending on requested complexity level (more complex combinations lead to larger feature spaces). 
These complex features are used in SISSO as a basis in materials space: the target property is expressed as a linear combination of complex features, but each complex feature may be a non-linear function of primary features.
Primary features themselves are also included in the basis.
Using complex features is a crucial step, because in general primary features may not correlate well with the target property, while their combination may represent a physical relation that describes target property very well.
For example, there is no physical reason why either electron affinity or ionization potential should correlate to electronic excitation energy, while the difference between them is directly physically related to it.

Here, we have produced an initial set with 20 primary features, see Table \ref{Tab:N1}.
This set is used to generate 1260 candidate features (complex plus primary).
The list of primary features includes the properties such as radii of the atoms in the compound, density, bulk and shear moduli, as well as the elasticity tensor components, elastic anisotropy, Poisson's ratio, Young's modulus, and more (see Table \ref{Tab:N1}).

\begin{table*}[t]
\begin{tabular}{lcc}
\hline \hline
Name                                               & Units & Abbreviation \\ \hline
Density                                            & $g/cm^{3}$ & D            \\
Voigt averaging of bulk modulus $B_V$                    & GPa   & $B_V$     \\
Reuss averaging of bulk modulus $B_R$                    & GPa   & $B_R$     \\
Voigt-Reuss-Hill averaging of bulk modulus $B_{VRH}$     & GPa   & $B_{VRH}$     \\
Voigt averaging of shear modulus $G_V$                    & GPa   & $G_V$     \\
Reuss averaging of shear modulus $G_R$                    & GPa   & $G_R$     \\
Voigt-Reuss-Hill averaging of shear modulus $G_{VRH}$     & GPa   & $G_{VRH}$     \\
Young's modulus                                    & GPa   & Y            \\
Fraction                                           &        & Fr           \\
Elastic   anisotropy                               &        & el           \\
Poisson’s   ratio &      &       $\sigma$           \\
Maximum atomic radius           & \AA     & $R_X$     \\
Minimum atomic radius           & \AA     & $R_N$     \\
Weighted atomic radius           & \AA     & $R_W$     \\
Maximum atomic weight           & a.m.u.     & $A_X$     \\
Minimum atomic weight          & a.m.u.     & $A_N$     \\
Weighted atomic weight           & a.m.u.     & $A_W$     \\
Maximum first ionization energy & eV    & $I_X$     \\ 
Minimum first ionization energy & eV    & $I_N$     \\ 
Weighted first ionization energy & eV    & $I_W$     \\ \hline \hline
\end{tabular}
\caption{Primary features used for construction of the descriptor}
\label{Tab:N1}
\end{table*}

All of the primary features utilized in this study were obtained from either the literature \cite{dean_langes_1999} or from the Materials Project database \cite{jain_materials_2013}.
The primary features were combined using the following set of operators\cite{ouyang_sisso_2018,Han2021-so,ouyang2019simultaneous,purcell2023recent,xu2022sisso,wei2022sisso}:
\begin{equation}
\label{eq:operators}
 \hat{H} \equiv \{+, -, *, /, ^{-1}, ^{2}, ^{3}, \sqrt, \sqrt[3], \exp, \log, |-| \} [\phi_1 , \phi_2], 
\end{equation}
\indent where $\phi_1$ and $\phi_2$ are primary features (in case of a unary operator, only one feature $\phi_1$ is considered).
The set of operators is applied recursively to generate features of increasing complexity.
Complexity level zero ($\Phi_0$) contains only primary features. $\Phi_1$ contains $\Phi_0$, features obtained by applying unary operators to all primary features, and binary combinations of primary features.
$\Phi_2$ contains $\Phi_1$ plus all new features obtained by unary and binary operations on $\Phi_1$. 
In this work, $\Phi_2$ is the highest considered level of complexity.

SISSO is used to both {\em select} the most important complex (or primary) features and find the model for the target property. 
Thus, SISSO automatically, as explained below, finds the most important primary features and the physically interpretable descriptor from data.
The number of selected features (descriptor dimension) depends on the required accuracy of training data fitting: the larger it is, the better is the fitting. 
However, larger number of complex features will eventually result in overfitting, which leads to worsening prediction accuracy for new materials not included in the training set.
The optimal number of complex features is determined by cross-validation.
In this work we have used 10-fold cross-validation (CV10). This consisted in randomly subdividing the data set in ten subsets and progressively using nine subsets for training the SISSO model and one subset for verification of the model.
The prediction (CV10) error is then evaluated as an average model error for the 10 verification subsets. 
As the number of complex features in the model (which is an input parameter) increases, the fitting error first decreases, but eventually starts to increase, indicating overfitting.
The dimension that yields minimum CV10 error is then used to find the best SISSO model using all training data.

\section{Results and Discussion}
\label{sec:res-dis}

\subsection{Development of the Descriptor}

\begin{figure*}[t]
    \centering
    \includegraphics[width=\textwidth]{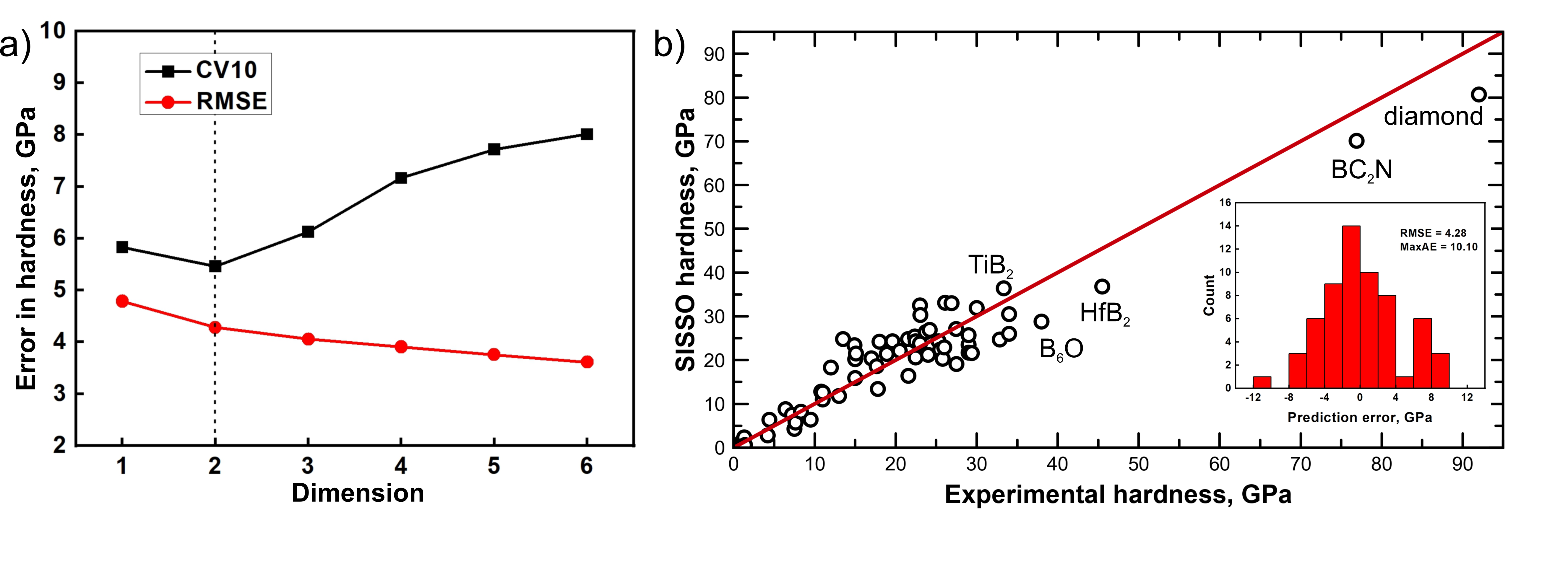}
    \caption{a) Root-mean square error (RMSE) for the SISSO model and the average RMSE of CV10. Dashed vertical line denotes the optimal descriptor dimension. b) The correlation between predicted hardness by 2D SISSO descriptor and experimental values of 61 compounds. The inset shows the distribution of RMSE and maximum absolute error (MaxAE) for the prediction of hardness 2D SISSO descriptor.}
    \label{fig:1}
\end{figure*}

We have collected a set of 635 compounds, selected from the Materials Project database \cite{jain_materials_2013}.
Materials without reliable experimental data for Vickers hardness (i.e., target property) were eliminated from the dataset, as those that were deemed unstable according to DFT calculations from aforementioned database.
This led to a total of 61 compounds for our training dataset, containing both hard materials (borides, carbides, nitrides, etc.) and comparatively soft ionic crystals and oxides (NaCl, Al$_{2}$O$_{3}$, etc.).
In order to access the values of such primary features and the properties for the training data sets, a request can be made via the GitHub (see link in the Appendix A).
The dataset for the target property (hardness) was created using information gathered from Zhang {\em et al.} \cite{zhang_determining_2021}.
The best found descriptors with different dimension (from 1D to 6D), see Eqs.~(A1)-(A6) are presented in the specific section in Appendix A. The CV10 error as a function of dimension is shown in Fig.~\ref{fig:1}a. As we can see, CV10 error increases from dimension larger than two, while the fitting root-mean square error (RMSE) reduces monotonically as the dimension of the descriptor increases (red curve in Fig.~\ref{fig:1}a).
CV10 is expected to increase in case of overfitting, which happens when the number of parameters, in this case the dimension of descriptor, is so high that the model learns random details and noise in the dataset, making it unable to correctly predict the property of unexplored materials.
Thus, according to CV10, the obtained optimal descriptor dimension is two.
This 2D descriptor bears a relatively complex analytical form:

\begin{eqnarray}
 {H_{predicted}^{SISSO} = 0.147 \cdot \frac{B_V}{\sigma \sqrt[3]{G_R}} - 1.136 \cdot \frac{B_R\log{R_X}}{A_W}-5.679}
 \label{eq:hardness}
\end{eqnarray}

\indent where $B_V$, $B_R$ are the values of bulk modulus calculated using Voigt and Reuss averaging methods, \cite{anderson_simplified_1963,hill_elastic_1952} respectively,
while $G_R$ is the shear modulus calculated using Reuss averaging method,
$\sigma$ is a Poisson's ratio, $A_W$ is the average atomic mass of the compound, and $R_X$ is the maximum atomic radius of the species in the compound.
The computed atomic radius and mass data were obtained from the Python library for materials analysis, Pymatgen \cite{ong_python_2013}.

The correlation between the optimal SISSO model and experimental hardness values is shown in Figure \ref{fig:1}b. The error distribution for hardness prediction using the optimal model with the 2D descriptor is shown as the inset to the Figure \ref{fig:1}b.
We attained a relatively low fitting RMSE of 4.28 GPa, as well as CV10 RMSE of 5.48 GPa for 2D model, with a maximum absolute error (MaxAE) of 10.1 GPa on the training set. The average relative error with respect to experimental hardness is \textcolor{red}{1.99} \%, which is small enough for a fast screening of promising candidates and lowest compared to what obtained with different previous models as Teter \cite{teter_computational_1998}, Chen \cite{chen_modeling_2011} and Mazhnik \cite{mazhnik_model_2019} that are respectively 21 \%, 14 \% and  16 \%.
The small CV10 error also indicates that the chosen set of primary features contains the important physical quantities that are necessary for describing the hardness.
Although including more advanced and therefore computationally more expensive primary features, e.g. surface or defect formation energies, may significantly improve the model, the reasonable predictive power of the current model indicates that these more advanced features can be approximately expressed through the employed primary features.

\subsection{Analysis of the Impact of Dimension of Descriptor }

Furthermore, to better understand the impact that each component of the two-dimensional descriptor has on the outcome, we calculated an importance score $IS$ for each term in Eq.~\ref{eq:hardness} towards the total error of our model.
This involved eliminating one component of the descriptor at a time and re-fitting the model with the remaining component.
The resulting one-dimensional derivative models are formulated as follows: 
\begin{equation}
{H_1 = a_1 \cdot \frac{B_R\log{R_X}}{A_W} + b_1}
 \label{eq:first}
\end{equation}
and
\begin{equation}
{H_2 = a_2 \cdot \frac{B_V}{\sigma \sqrt[3]{G_R}} + b_2}
 \label{eq:second}
\end{equation}
Coefficients $a_1$, $b_1$, and $a_2$, $b_2$ were fitted separately for $H_1$ and $H_2$ by minimizing RMSE, and are equal to $a_1$ = 15.384, $b_1$ = 0, and $a_2$ = 0.1485, $b_2$ = -7.2.

The $IS$ is calculated by using the RMSE and MaxAE values for $H_1$ and $ H_2$, respectively, for our dataset as follows:
\begin{equation}
 {IS_i^{\rm RMSE} = 1-\frac{{\rm RMSE}(H_{predicted})}{{
\rm RMSE}(H_i)}}
     \label{eq:rmse}
\end{equation}
\begin{equation}
{IS_i^{\rm MaxAE} = 1-\frac{{\rm MaxAE}(H_{predicted})}{{\rm MaxAE}(H_i)}}
    \label{eq:ae}
\end{equation}

Calculated importance scores based on RMSE and MaxAE are respectively 0.49 and 0.52 for $IS_{1}$,and 0.07 and 0.06 for $IS_{2}$. 
Thus, the first descriptor component in Eq.~\ref{eq:hardness} plays a more significant role in hardness according to our model.
However, including both descriptor components in the SISSO model reduces the RMSE and MaxAE errors by 6-7\%. 
The RMSE on the shared dataset is 5.2 GPa for $H_1$ and 9.3 GPa for $H_2$.
The use of both $H_1$ and $H_2$ results in a lower error value of 4.28 GPa, highlighting the importance of the 2D descriptor in comparison to the 1D counterpart. 

The obtained 2D model was used to perform high-throughput screening of hard and superhard materials belonging to binary, ternary, and quaternary transition metal borides, carbides, and nitrides.
The required crystal structures of experimentally known and hypothetical structures were extracted using the Materials Project database \cite{jain_materials_2013}.
In total, 635 structures were gathered for the selected classes of materials.
For each structure, we have extracted the necessary properties for the developed model, including bulk and shear moduli, Poisson's ratio, and the averaged atomic mass of each compound. 
The maximum radius of the atoms in the compound was determined using the Pymatgen library \cite{ong_python_2013}.

\begin{figure*}[t]
    \centering
    \includegraphics[width=\textwidth]{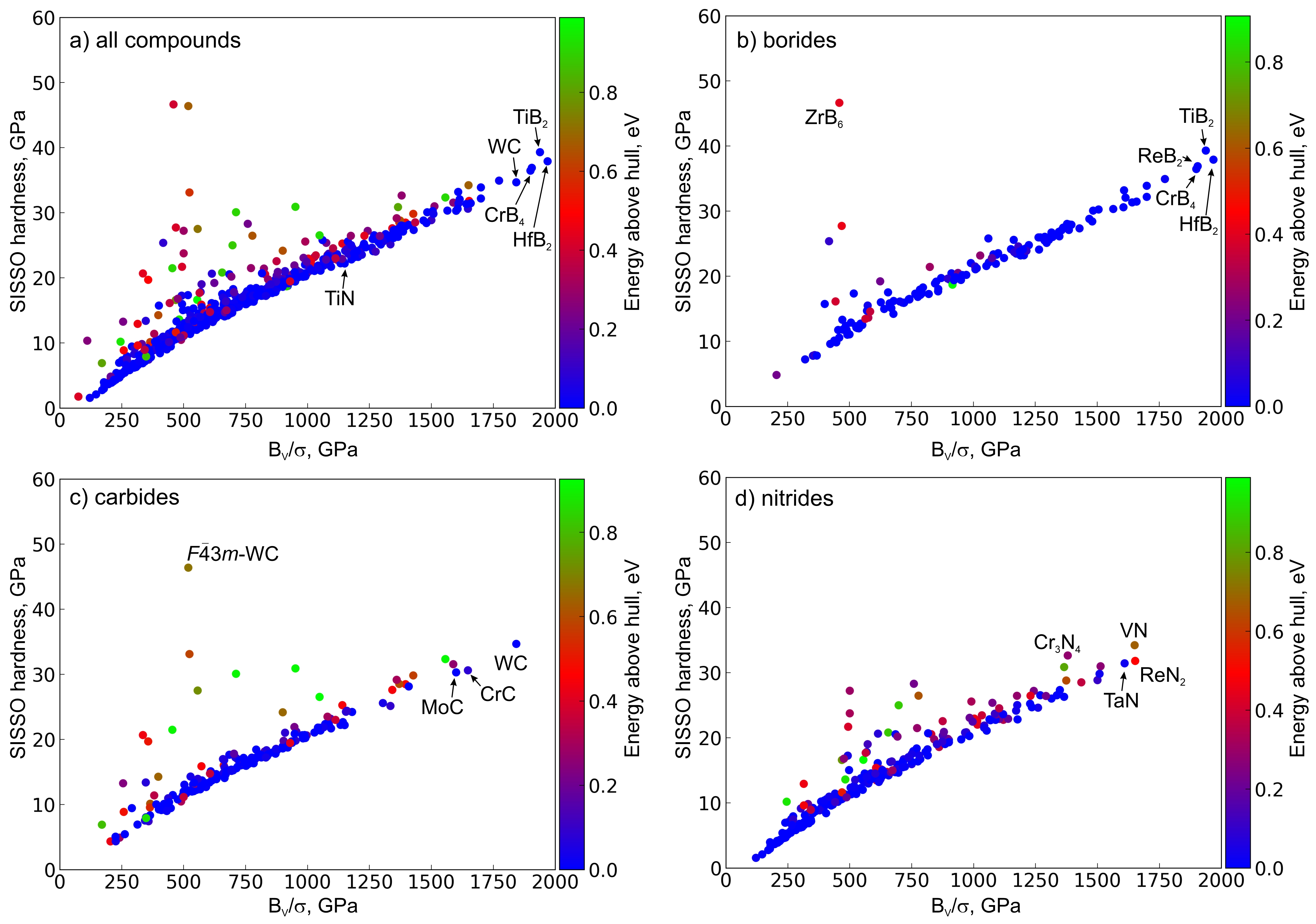}
    \caption{a) The SISSO $H_{V}$ model predictions are plotted against $B_v/\sigma$ for considered 635 inorganic compounds. Specific classes of materials are also shown, including b) borides, c) carbides, and d) nitrides. Colorbar shows the energy of formation above the convex hull denoting stability of each structure.}
    \label{fig:distr}
\end{figure*}

To analyse the collected data, we have constructed the correlation plot displaying the relationships between SISSO Vickers hardness, bulk modulus, Poisson's ratio, and shear modulus for 635 inorganic compounds.
This excludes diamond, borocarbides, carbonitrides, and layered compounds, as shown in Figure \ref{fig:distr}a.
The color scale of the points indicated the energy above the convex hull to represent the (meta)stability of each compound. 
A clear trend in increasing hardness with higher  $B_v/\sigma$ values is visible. 
There are outliers which show high hardness and quite low shear modulus, together with a low $B_v/\sigma$ value which contradicts the general trend. 
These outliers correspond to metastable structures (see red and green points in Figure \ref{fig:distr}a).
Despite the denominator of Eq.~\ref{eq:hardness} containing $G_R$, the correlation between $B_V$ and $G_R$ (Figure B.1 in Appendix B) results in an overall increase in hardness as shear modulus increases.
The correlation between $B_V$ and $G_R$ for stable structures is further enhanced (see Figure B.2 in the Appendix B).
Moreover, compounds conforming to the general trend exhibit the Pugh's ratio  (i.e., G/B) ranging from 0.5 to 0.8, as shown in Figure B.1 and B.2 in Appendix B, it is indicative of how brittle the material becomes or not.
This supports non-linear relationship between hardness and other properties, emphasising the significance of accounting for this non-linearity to identify hard materials.

In Figure \ref{fig:distr}a, well-known hard and superhard compounds for a total of 635 are identified as reference points.
This aids in understanding the location of other compounds in relation to them.
The highest values of hardness belong to boride and carbide compounds (see Figure \ref{fig:distr}b,c). 

Among selected borides (Figure \ref{fig:distr}b), ZrB$_6$ (mp-1001788), a metastable compound located 0.4 eV/atom above the convex hull (according to data from the Materials Project) and with a predicted SISSO hardness of 46 GPa, can be highlighted. 
ZrB$_6$ has a crystal structure similar to that of calcium hexaboride, consisting of a 3D boron cage, which contributes to its high bulk modulus and hardness.
The influence of the boron cage on the mechanical and elastic properties of borides has previously been demonstrated for hafnium borides \cite{xie_stable_2019}.
It should be noted that, while such a crystal type is typical for borides of rare-earth elements, it is an unusual metastable structure for transition metals, with an extremely low Reuss-averaged shear modulus of 2 GPa and a low $B_v/\sigma$ of 500 GPa (the Poisson's ratio is 0.39). 
However, this discovery suggests a promising way to enhance the hardness of rare-earth borides by incorporating transition metals as substitutes within the crystal structure.
Also, high hardness is predicted for well-known superhard compounds, namely TiB$_2$, ReB$_2$, HfB$_2$, and CrB$_4$ (see Figure~\ref{fig:distr}b).

Among the carbides, cubic polymorphic modification of tungsten carbide (WC) with $F\bar{4}$3$m$ space group (see Figure \ref{fig:distr}c) has the highest hardness of 46 GPa.
WC (mp-1008635) has a zincblende structure where each tungsten atom forms corner-sharing WC$_4$ tetrahedra with four equivalent carbon atoms. 
The structure has a bulk modulus of 249 GPa and a shear modulus of 3 GPa, resulting in a very high Poisson's ratio of 0.48.
Despite its high hardness, this structure is deemed unstable with an energy of formation 0.67 eV/atom above the convex hull (according to data from the Materials Project).
The well-known hexagonal modification of WC has an SISSO hardness of 35 GPa with bulk and shear moduli equal to 387 and 276 GPa respectively. 
Predicted values are in good agreement with experimental data and those obtained by other models \cite{kvashnin_computational_2019}.
Hexagonal WC has the highest $B_v/\sigma$ ratio compared to the other considered carbides at a value of 1842 GPa. 
Two other structures with comparable mechanical characteristics to WC are CrC (mp-1018050) and MoC (mp-2305) as shown in Figure \ref{fig:distr}c.
Both of these have a hexagonal $P\bar{6}m$2 space group, the same as in hexagonal WC.
Each metal atom in the structure forms bonds with six equivalent carbons to create a mixture of distorted face, edge, and corner-sharing MeC$_6$ pentagonal pyramids.
It is predicted that the SISSO hardness of both CrC and MoC is approximately 30 GPa. 
The bulk modulus of both structures is roughly 350 GPa, whereas the shear modulus is about 240 GPa.
CrC proves metastable with an energy of formation 80 meV/atom above the convex hull, whereas MoC is stable and holds a calculated energy of formation of only 1 meV/atom above the convex hull. 

The hardest found compounds among nitrides are VN, TaN, and ReN$_2$, see Figure \ref{fig:distr}d.
VN (mp-1002105) belongs to the $Pm\bar{3}m$ space group and is located 0.68 eV/atom above the convex hull.
The SISSO model predicts its hardness to be 34 GPa with $B_v/\sigma = 1650$ GPa (while Poisson's ratio is 0.16).
TaN (mp-1009831), which has a SISSO hardness of 31 GPa is isostructural to well-known WC structure and belongs to the $P\bar{3}m$2 space groups. 
It has a Poisson's ratio of 0.21 and $B_v/\sigma = 1610$ GPa, as shown in Figure \ref{fig:distr}d.
Renium dinitride (mp-1019055) is located 0.49 eV/atom above the convex hull and predicted to have a hardness of 32 GPa with $B_v/\sigma = 1650$ GPa.

Another interesting nitride material is Cr$_3$N$_4$ (mp-1014460), see Figure \ref{fig:distr}d.
The material has a $Pm\bar{3}m$ space group and can be depicted as a rocksalt structure with a missing atom in the 4\textit{a} Wyckoff position which results in fractional composition. 
Its predicted SISSO hardness is 33 GPa, and it has a low Poisson's ratio of 0.1, leading to a high $B_v/\sigma$ value of 1380 GPa.

\begin{figure*}[t]
    \centering
    \includegraphics[width=\textwidth]{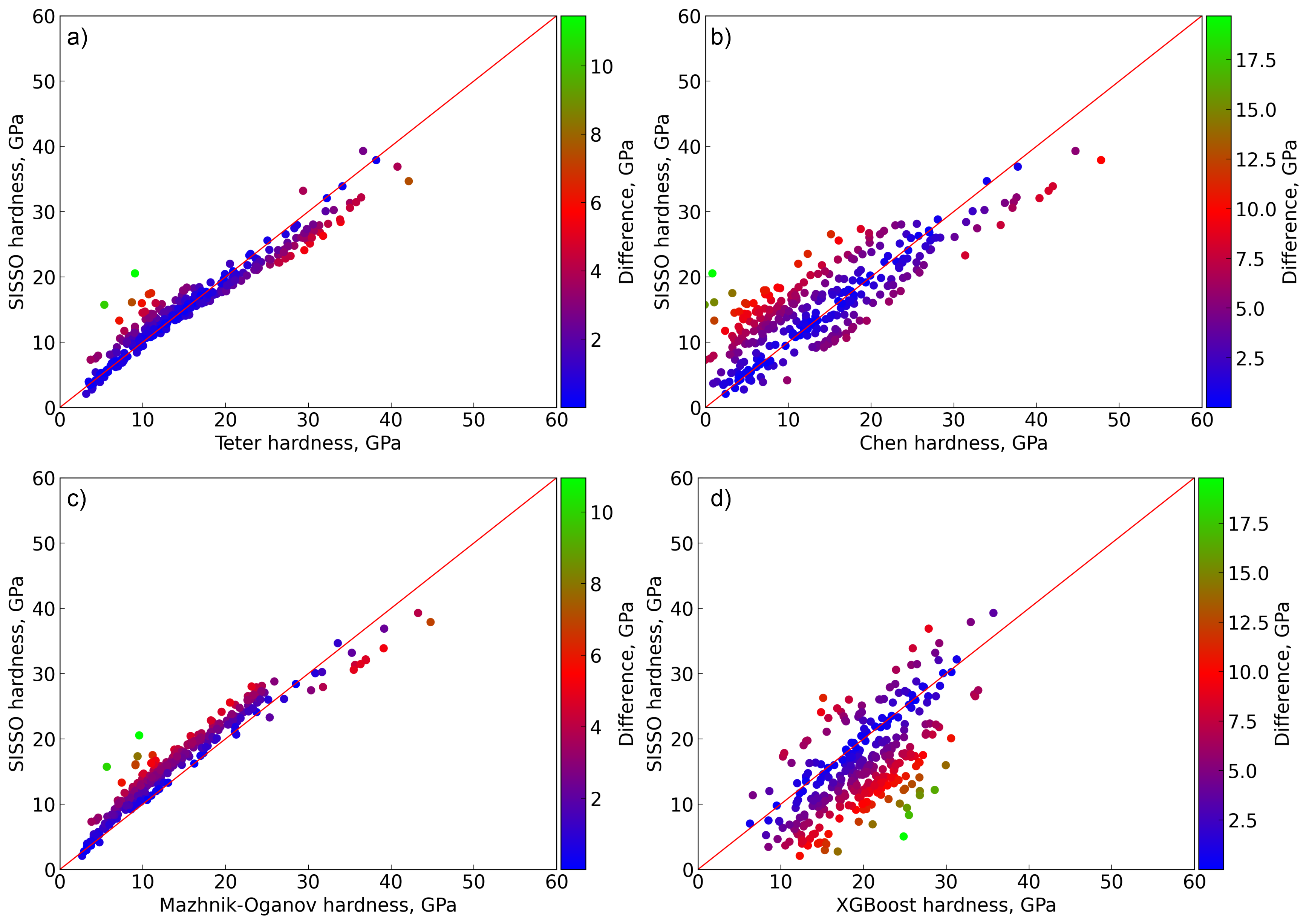}
    \caption{Correlations between SISSO hardness and a) Teter \cite{teter_computational_1998}, b) Chen \cite{chen_modeling_2011}, c) Mazhnik-Oganov \cite{mazhnik_model_2019}, d) XGBoost \cite{zhang_finding_2021} models for considered stable structures. Colorbar shows the difference between two sets of data.}
    \label{fig:modelsvssisso}
\end{figure*}

Furthermore, we compare our SISSO hardness model with other machine learning and empirical models, we have used Teter's \cite{teter_computational_1998}, Chen's \cite{chen_modeling_2011}, Mazhnik-Oganov's \cite{mazhnik_model_2019}, and XGBoost \cite{zhang_finding_2021} models to predict the hardness of structures in the created dataset.
The Figure \ref{fig:modelsvssisso} portrays their correlations with the SISSO model for stable structures lied on the convex hull.
The color scale indicated variations between the SISSO and the considered reference model. 
Our model yields a good agreement of predicted hardness values with the Teter model, as shown in  Figure \ref{fig:modelsvssisso}a. 
The greatest difference between predictions was 12 GPa for hexagonal NaBPt$_3$ (mp-28614), and the next greatest was 20 GPa for zincblende FeN (mp-6988). 
The largest deviation between the SISSO model and Chen's model was found to be 15.5 GPa for NaBPt$_3$ (see Figure \ref{fig:modelsvssisso}b). 
The SISSO hardness of this compound is 21.2 GPa, whereas Chen's hardness is only around 5 GPa. 
This significant variation could be attributed to the highly anisotropic structure of NaBPt$_3$, resulting in a difference of 33 GPa between Reuss- and Voigt-averaged shear moduli according to the Materials Project. 
In our model, we use Reuss averaging, resulting in higher hardness than Chen's model, which uses the Voigt-Reuss-Hill averaged shear modulus.
The latter is lower compared to the Reuss averaged value for NaBPt$_3$.
Predictions of our model align well with the recent Mazhnik-Oganov model\cite{mazhnik_model_2019} as shown in Figure \ref{fig:modelsvssisso}c, except for NaBPt$_3$ and FeN, where the differences are similar to Teter's model.

The use of machine-learning XGBoost model for predicting hardness was innovative and highly efficient \cite{zhang_finding_2021}. 
We trained the same XGBoost model as was used in Ref. \cite{zhang_finding_2021} on our training set, and predicted hardness for all the considered compounds. 
First, we performed the 10-fold cross-validation using the same techniques and dataset as for SISSO training, that is, we divided the dataset into 10 subsets and trained the XGBoost model using 9 of those subsets. 
The CV10 error was calculated as the mean value of the test RMSE acquired for each of the ten subsets, and equated to 7.8 GPa, which is approximately twice as high as the CV10 error for SISSO. 
The distribution of errors for the XGBoost model for CV10 is shown in the Appendix B (Figure B.3).
The correlations between the XGBoost model and the SISSO model is shown in Figure \ref{fig:modelsvssisso}d.
Numerous structures have a hardness disparity ranging from 12 to 17 GPa.
Most of these structures comprise rare-earth metal carbides, specifically Y$_2$C (mp-1334), Sc$_4$C$_3$ (mp-15661), Y$_4$C$_5$ (mp-9459), Y$_2$ReC$_2$ (mp-21003).
Such significant differences in the hardness predicted by the XGBoost and SISSO models for our compounds can be attributed to the fitting hyperparameters of the XGBoost algorithm, which need to be redefined before training on the new training set.
When considering only transition-metal borides, carbides, and nitrides, much lower differences can be obtained between XGBoost and SISSO, as XGBoost more accurately describes these classes of compounds (see Figure B.5 in the Appendix B). 

Our findings demonstrate that SISSO identified a physical significance of the $B_v/\sigma$ ratio for hardness, enabling one to quickly estimate the hardness of a compound across a diverse range of chemical compositions and crystal structures.
A greater $B_v/\sigma$ ratio corresponds to heightened hardness.

%%%%%%%%%%%%%%%%%%%%%%%%%%%
\section{Conclusion}
\label{sec:conclusion}

In conclusion, our study successfully identified a physical descriptor for Vickers hardness through a novel application of the SISSO artificial-intelligence algorithm. By leveraging a symbolic-regression approach based on compressed sensing, we have derived a non-linear function correlating microscopic properties to macroscale hardness for a wide range of materials. The key contributors to this descriptor are the Voigt-averaged bulk modulus, Poisson's ratio, and Reuss-averaged shear modulus, reflecting the intricate relationship between these properties and material hardness. The model, validated against experimental values for a diverse set of transition-metal compounds, demonstrates significant predictive power. High-throughput screening of 635 candidate materials using this descriptor reveals promising pathways for enhancing material hardness, particularly through the incorporation of harder, metastable structures such as VN, TaN, ReN$_2$, Cr$_3$N$_4$, and ZrB$_6$.

Broadly, these findings underscore the transformative potential of artificial intelligence in materials science. By bridging the gap between atomic-level properties and macroscale characteristics, such approaches can accelerate the discovery and design of advanced materials with tailored properties, driving innovation across various industrial sectors. This research not only provides a robust tool for predicting material hardness, but also highlights the importance of interdisciplinary methodologies in solving complex materials challenges.

\section*{Acknowledgement}
Ch.T was supported by the Norwegian Research Council through a Centre of Excellence grant (Hylleraas Centre 262695), a FRIPRO grant (ReMRChem 324590).
T.A. and B.I.Y. acknowledge the Taif University Research
Support Project (TURSPHC2024/1, Saudi Arabia).

 \bibliographystyle{elsarticle-num} 
 \bibliography{cas-refs}

\appendix
\section{Predicted descriptors}
All the data about datasets are available via the github link by request: 
\url{github.com/AlexanderKvashnin/SISSO_hardness.git}. 
There is a list of predicted descriptors by SISSO used for calculations the RMSE and CV10 in Figure 1a.

\begin{align}
 {H^{1D} = 0.182 \cdot \frac{B_R}{\sigma \sqrt[3]{Y}} - 6.191}
\end{align}

\begin{align}
 {H^{2D} = 0.147 \cdot \frac{B_V}{\sigma \sqrt[3]{G_R}} - 1.136 \cdot \frac{B_R\log{R_X}}{A_W}-5.679}
\end{align}

\begin{align}
 H^{3D} & = 0.659 \cdot \frac{B_R}{\sigma \sqrt[3]{Y}} - 1.405 \cdot \frac{G_V}{A_W} \cdot \log{R_X} \nonumber \\ 
 & - 0.042 \cdot \frac{Fr}{R_N \log{el}} - 12.221
\end{align}

\begin{align}
 H^{4D} & = 0.677 \cdot \frac{B_R}{\sigma \sqrt[3]{Y}} - 0.133 \cdot \frac{Y}{D} \cdot \log{R_X} \nonumber \\
 & + 0.041 \cdot \frac{Fr}{R_N \log{el}} - 13.228 \cdot \frac{I_W}{I_X \sqrt{R_W}} - 1.471
\end{align}

\begin{align}
 H^{5D} & = 0.155 \cdot \frac{B_R}{\sigma \sqrt[3]{G_V}} - 0.353 \cdot \frac{G_V}{D} \cdot \log{R_X} + 0.054 \cdot \frac{Fr}{R_W \log{el}} \nonumber \\
  & - 1027 \cdot \frac{|B_V - G_R|}{\exp{A_N}} +3.190 \cdot \frac{R_W}{el |B_R - G_V|} -5.873
\end{align}

\begin{align}
 H^{6D} & = 0.177 \cdot \frac{B_R}{\sigma \sqrt[3]{G_V}} - 41.972 \cdot \frac{\log{R_X}}{A_W} \cdot \sigma + 0.046 \cdot \frac{G_R}{R_N \log{el}} \nonumber  \\ 
  & - 1175 \cdot \frac{|B_R - G_R|}{\exp{A_N}} + 0.047 \cdot \frac{D^3}{|B_V - G_V|} \nonumber \\
  & - 0.963 \cdot \frac{A_X}{A_W} \cdot \sqrt{A_N} + 3.815
\end{align}

\clearpage
\section{Additional data}
\setcounter{figure}{0}
\begin{figure}[ht]
    \centering
    \includegraphics[width=\textwidth]{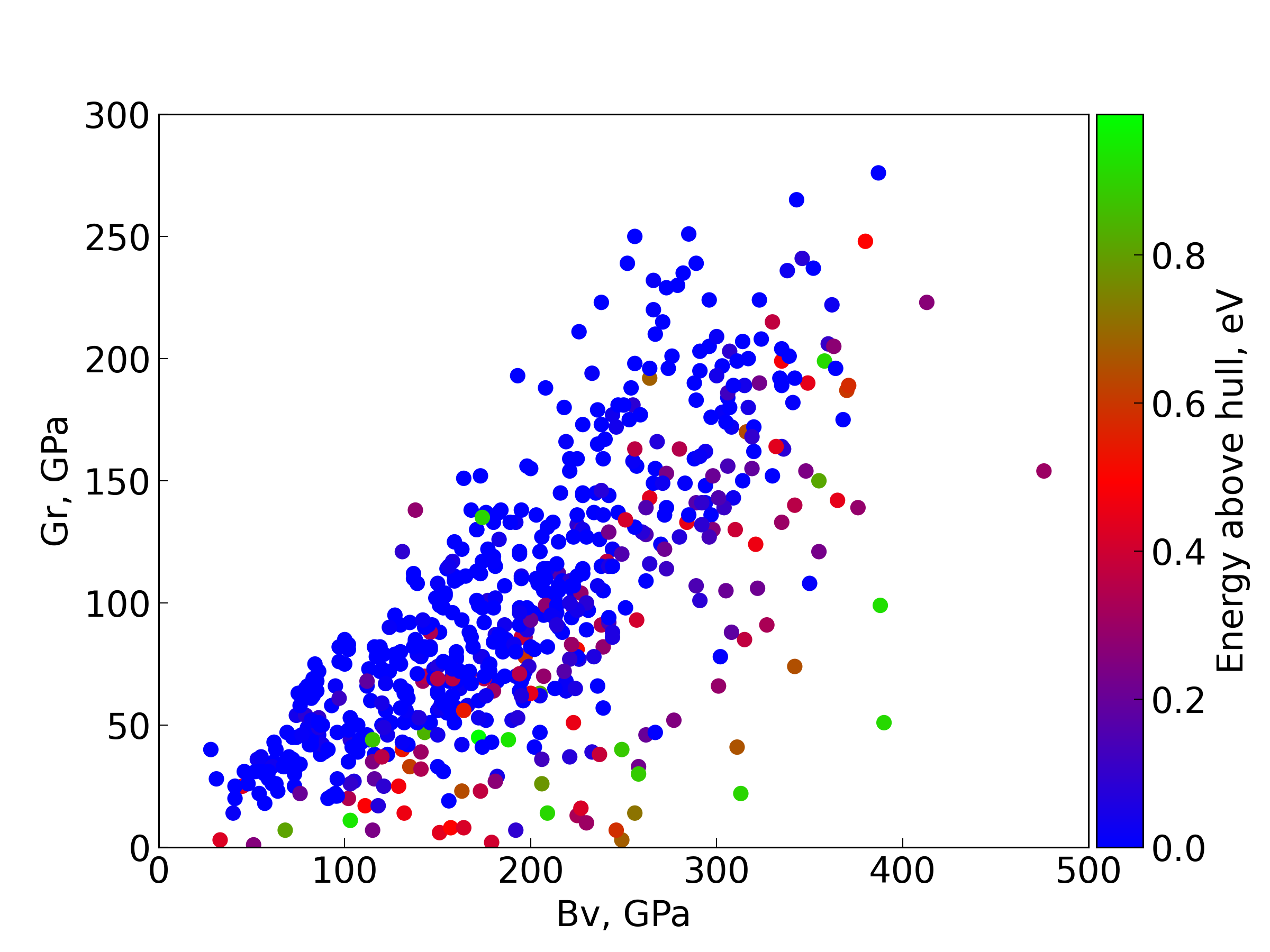}
    \caption{Correlation between Voigt-averaged bulk modulus and Reuss-averaged shear modulus of stable and metastable structures among borides, carbides, and nitrides. Colorbar shows the energy of formation above the convex hull denoting stability of each structure.}
\end{figure}

\begin{figure}[ht]
    \centering
    \includegraphics[width=\textwidth]{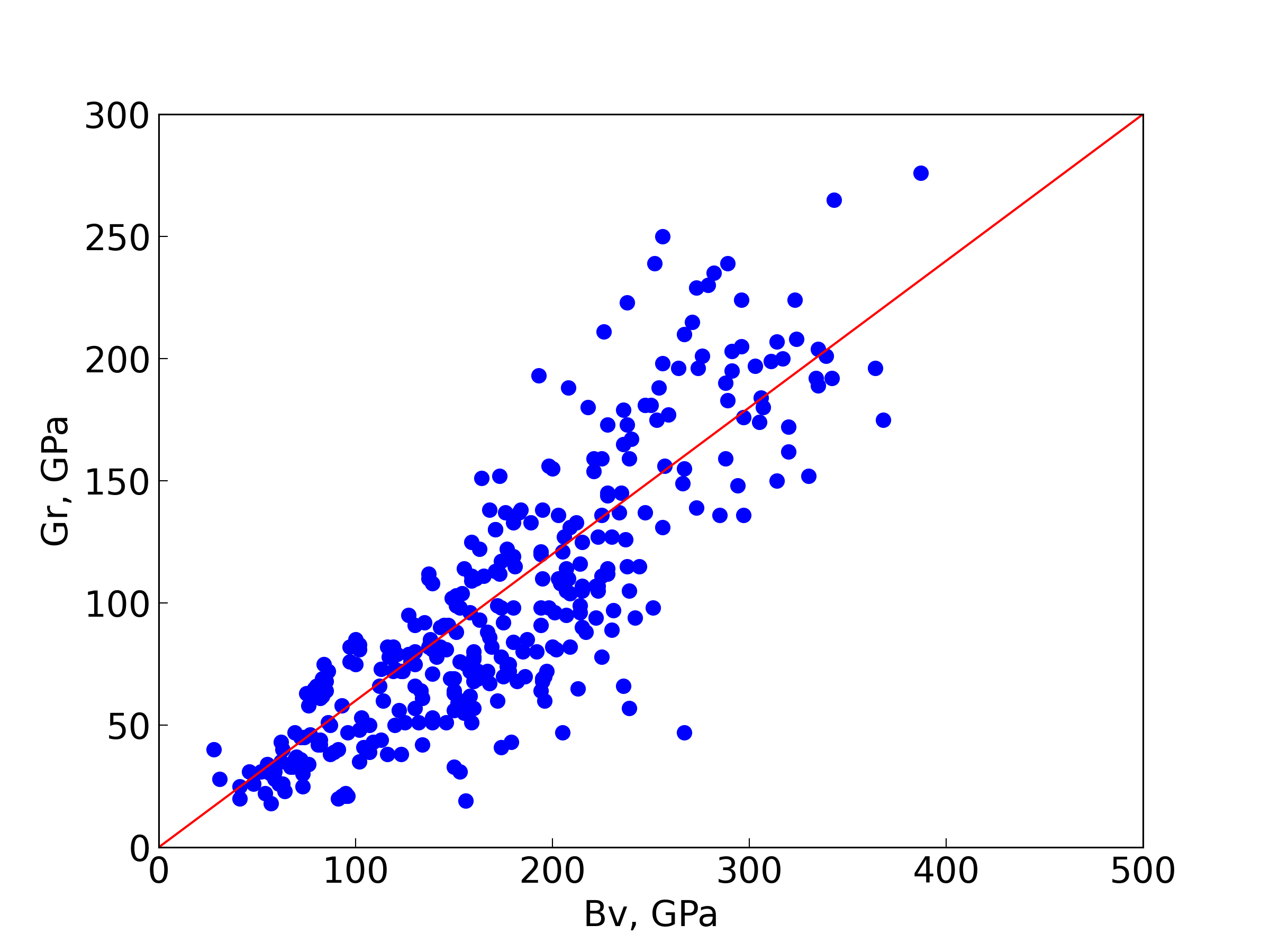}
    \caption{Correlation between Voigt-averaged bulk modulus and Reuss-averaged shear modulus of only stable structures among borides, carbides, and nitrides.}
\end{figure}

%\begin{figure}[ht]
%    \centering
%    \includegraphics[width=\textwidth]{SI/Hardness_vs_Bv-sigma_pough.png}
%    \caption{Correlation between SISSO hardness and $B_v/\sigma$ ratio of stable and metastable structures among borides, carbides, and nitrides. Colorbar shows the Pough ratio.}
%\end{figure}

\begin{figure}[ht]
    \centering
    \includegraphics[width=\textwidth]{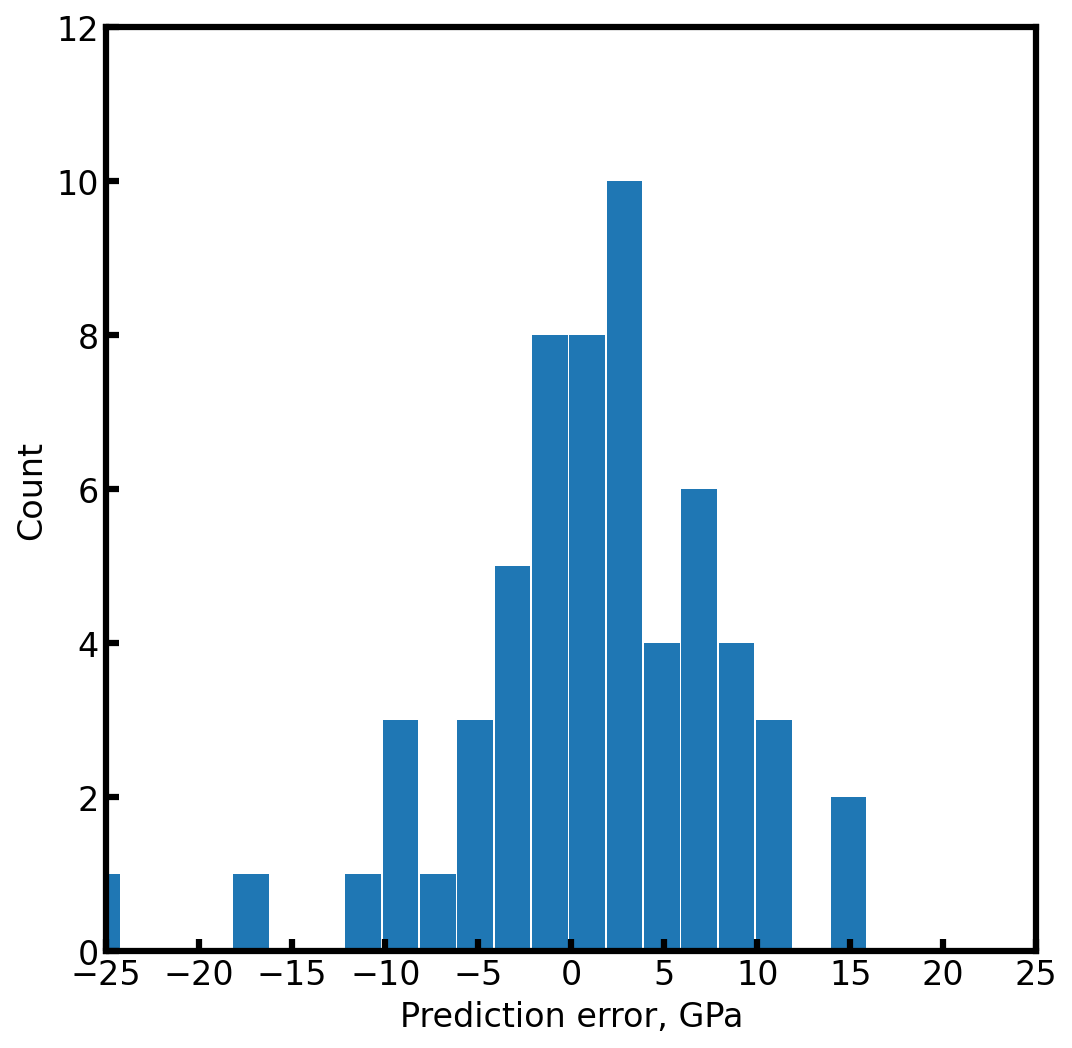}
    \caption{Distribution of CV10 errors for XGBoost model. Maximum absolute error is 25.6 GPa, RMSE is 7.8 GPa.}
\end{figure}

%\begin{figure}[ht]
%    \centering
%    \includegraphics[width=\textwidth]{SI/XGBoost_vs_SISSO_stable_carb_bor_nitr.png}
%    \caption{Correlation between SISSO hardness and XGBoost \cite{zhang_finding_2021} model for considered stable carbides, borides and nitrides only. Colorbar shows the difference between two sets of data.}
%\end{figure}

\end{document}